\documentclass[twocolumn,prd,nofootinbib,aps,prl,floats,floatfix,amsmath,amssymb,longbibliography,secnumarab
ic]{revtex4-1} %
\usepackage[english]{babel}

\usepackage[letterpaper,top=2cm,bottom=2cm,left=3cm,right=3cm,marginparwidth=1.75cm]{geometry}

\usepackage{amsmath}
\usepackage{graphicx}
\usepackage[colorlinks=true, allcolors=blue]{hyperref}
\newcommand{\be}{\begin{equation}}
\newcommand{\ee}{\end{equation}}

\begin{document}

\title{Comment on ``Excluding Primordial Black Holes as Dark Matter Based on Solar System Ephemeris''}
\author{James M.\ Cline}
\email{jcline@physics.mcgill.ca}
\affiliation{McGill University Department of Physics \& Trottier Space Institute, 3600 Rue University, Montr\'eal, QC, H3A 2T8, Canada}
\affiliation{CERN, Theoretical Physics Department, Geneva, Switzerland}

\begin{abstract}
It was recently claimed (\url{https://arxiv.org/pdf/2408.10799}) that
solar system ephemeris data can exclude primordial black hole (PBH) dark matter in the mass range $10^{18}-10^{22}$\,g.  We show that this conclusion is based on an implausible, implicit assumption; namely the uncertainty on the solar system mass within 50 au is as small as the uncertainty on the mass of the sun.  Correcting for this error, we find that ephemeris data can only constrain PBH's with mass below $10^{16}$\,g, which is already excluded by constraints on their evaporation via Hawking radiation.   Correcting a further error concerning the time-averaged rate of such fluctuations nullifies even this weaker constraint. 
\end{abstract}

\maketitle

Ref.\ \cite{Loeb:2024tcc} recently argued that if primordial black holes (PBHs) in the mass range $10^{18}-10^{22}$\,g constitute the dark matter of the Universe, they would give rise to Poisson fluctuations in the total mass of the solar system contained within some assumed fiducial volume, and that observational limits on such fluctuations can be used to exclude this PBH mass range.  It was claimed that fluctuations due to crossing PBHs would show up as an effective time rate of change of the solar mass $M_\odot$, by their perturbation of planetary orbits.
These orbits have been monitored over more than 50 years, and the search for deviations was used by Ref.\ \cite{Pitjeva:2021hnc} to set strong limits on the time variation of $GM_\odot$, at the level of a few parts in $10^{14}$ per year.

In Ref.\ \cite{Loeb:2024tcc}, Loeb assumed the relevant volume
is the region within 50 au, the orbital radius of Pluto.
However, Ref.\ \cite{Pitjev:2013sfa} shows that the ephemeris data are vastly dominated (94\,\%) by measurements of the inner planets.
Therefore a fiducial radius of 1.5 au (encompassing Mars) would be more appropriate for the data at hand.  To have at least one PBH on average within this volume, in order to cause a fluctuation, it should have a mass no greater than $3\times 10^{16}\,$g, assuming a monochromatic spectrum of PBHs making up the total dark matter density.  Such a light PBH as dark matter is already ruled out by constraints on its Hawking radiation evaporation products,
namely photons and $e^+$-$e^-$ pairs \cite{Carr:2020gox}.

However, even this weakened limit on the PBH mass cannot be a valid
inference from the ephemeris data.  The limit on $d/dt(GM_\odot)$
derived by Ref.\ \cite{Pitjeva:2021hnc} relies upon a long duration,
$\sim 50$\,y of data taking, in order to be so stringent.
In contrast, the crossing time of the 1.5\,au fiducial volume by a PBH is less than a few weeks, hence any perturbation produced on planetary orbits would average to zero during the observation time.
Moreover, the direct limit on the dark matter density in the solar system from ephemeris data is 14,000 times weaker than the accepted value $0.4\,$GeV/cm$^3$ \cite{Pitjev:2013sfa}, underscoring the implausibility of these data being sensitive to such a weak perturbation as a PBH crossing.

Not discussed in Ref.\ \cite{Loeb:2024tcc} is the possibility that a close encounter of a PBH could strongly perturb an inner planet.  An order of magnitude estimate shows that this is highly unlikely.  The ephemeris bound on the integrated perturbation to the gravitational force on a planet of mass $m$ is of order
\be
    \delta F \sim {G \dot M_\odot T m\over R^2}\,,
\ee
where $\dot M_\odot$ is the limit on the time variation of the solar mass,  $T$ is the observing time, and $R\sim 1\,$au is the orbital radius.  The maximum force exerted by a
transiting PBH is $F = G M_{bh} m/b^2$, where $b$ is the distance of closest approach.  Thus one needs 
\be
    b\lesssim R\left(M_{bh}\over \dot M T\right)^{1/2} \sim 0.01\,{\rm au}
\ee
for $M_{bh}=10^{18}$\,g, to have an observable effect.  On the other hand, Ref.\ \cite{Loeb:2024tcc} computes the rate of such crossings to be
\be
    \Gamma \sim 1000 \left(b\over 50\,{\rm au}\right)^2 {\rm yr}^{-1}
     < 4\times 10^{-5}\, {\rm yr}^{-1}\,.
\ee

{\bf Acknowledgment.}  I thank Gabriele Franciolini for helpful discussions.

\bibliographystyle{utphys}
\bibliography{sample}

\end{document}